\documentclass[english,tightenlines,showkeys]{article}
\usepackage[T1]{fontenc}
\usepackage[latin9]{inputenc}
\usepackage{geometry,times}
\usepackage{xcolor}
\usepackage{booktabs}
\usepackage{mathrsfs}
\usepackage{amsmath}
\usepackage{amssymb}
\usepackage{graphicx}
\PassOptionsToPackage{normalem}{ulem}
\usepackage{ulem}
\usepackage{textcomp}
\usepackage{subfigure,array}
\usepackage[sort&compress,numbers]{natbib}
\usepackage{nohyperref}

\makeatletter

\date{}

\makeatother

\usepackage{babel}
\begin{document}
\title{Manipulating nonclassicality via quantum state engineering processes:
Vacuum filtration and single photon addition}
\author{Priya Malpani$^{1}$, Nasir Alam$^{2}$, Kishore Thapliyal$^{2,3}$,
Anirban Pathak$^{2}$
V. Narayanan$^{1}$, Subhashish Banerjee$^{1}$\\
{\small $^{1}$Indian Institute of Technology Jodhpur, Jodhpur 342011, India}\\
{\small $^{2}$Jaypee Institute of Information Technology, A-10, Sector-62,
Noida, UP-201307, India}\\
{\small $^{3}$RCPTM, Joint Laboratory of Optics of Palacky University
and Institute of Physics of Academy of Science of the Czech Republic,}\\
{\small
Faculty of Science, Palacky University, 17. listopadu 12, 771 46 Olomouc,
Czech Republic}}
\maketitle
\begin{abstract}
The effect of two quantum state engineering processes that can be
used to burn hole at vacuum in the photon number distribution of quantum
states of radiation field are compared using various witnesses of
lower- and higher-order nonclassicality as well as a measure of nonclassicality.
Specifically, the witnesses of nonclassical properties due to the
effect of vacuum state filtration and a single photon addition on
an even coherent state, binomial state and Kerr state are investigated
using the criteria of lower- and higher-order antibunching, 
squeezing and  sub-Poissonian photon statistics. Further,
the amount of nonclassicality present in these engineered
quantum states is quantified and analyzed by using an entanglement potential based on linear entropy.  It is observed
that all the quantum states studied here are highly nonclassical, and on many
occasions the hole burning processes are found to introduce/enhance
nonclassical features. However, it is not true in general. The investigation has further revealed that despite the fact that a hole at
vacuum implies a maximally nonclassical state (as far as Lee's nonclassical
depth is used as the quantitative measure of nonclassicality). However, any
particular process of hole burning at vacuum does not ensure  the existence of a particular nonclassical feature. Specifically,
 lower- and higher-order squeezing are not observed for photon added even coherent state and vacuum filtered even coherent state. 
\end{abstract}

\section{Introduction }

The art of generating and manipulating quantum states
as per need is referred to as the ``quantum state engineering''
\cite{dakna1998quantum,sperling2014quantum,vogel1993quantum,miranowicz2004dissipation,dell2006multiphoton}.
This relatively new area of research has drawn much attention of
the scientific community because of its success in experimentally
producing various quantum states \cite{zavatta2004quantum,torres2003preparation,rauschenbeutel2000step,gao2010experimental,lu2007experimental}
having nonclassical properties and applications in realizing quantum
information processing tasks, like quantum key distribution \cite{bennett1984quantum}
and quantum teleportation \cite{brassard1998teleportation,chen2015bidirectional}.
Engineered quantum states, such as cat states,
Fock state and superposition of Fock states, are known to play a crucial
role in performing fundamental tests of quantum mechanics and in establishing
quantum supremacy in the context of quantum computation and communication
(\cite{kues2017chip} and references therein). 

In fact, to establish quantum supremacy or to perform a fundamental
test of quantum mechanics, we would require a state having some features
that would not be present in any classical state. Such a state having
no classical analogue is referred to as the nonclassical state and
is characterized by the negative values of Glauber-Sudarshan $P$
function \cite{sudarshan1963equivalence,glauber1963coherent}.
Frequently used examples of nonclassical states include squeezed, antibunched, entangled, steered, and Bell nonlocal
states; and the relevance of the states having these nonclassical features
have already been established in various domains of physics. For example,
squeezed state has been used in the successful experimental detection
of gravitational wave in LIGO \cite{aasi2013enhanced}, antibunching 
is used in characterizing single photon sources \cite{pathak2010recent}
required in quantum cryptography, and entangled states are useful in almost
all sub-fields of quantum information processing  \cite{pathak2013elements,pathak2017classical}.  In this article, we focus on two quantum state engineering operations ({namely, hole burning at vacuum by filtration  and single photon addition})
to evaluate their role in inducing/enhancing the nonclassicality in the engineered output state.

 To introduce the idea of these quantum state engineering operations, we can write the photon number distribution of an arbitrary quantum state in terms of Glauber-Sudarshan $P\left(\alpha\right)$ function as 
\begin{equation}
p_{n}=\int P\left(\alpha\right) \left|\langle n |\alpha\rangle\right|^2 d^2\alpha.\label{eq:pnd}
\end{equation}
If $p_{n}$ vanishes for a particular value of Fock state parameter $n$, we refer to
that as a ``hole'' or a hole in the photon number distribution at position $n$
\cite{escher2004controlled}. Notice that $p_{n}=0$ reveals that $P\left(\alpha\right)<0$ for some $\alpha$, which is the signature of nonclassicality. Thus, the existence of a hole in the photon number
distribution implies that the corresponding state is nonclassical,
and corresponding technique of quantum state engineering to create hole
is called hole burning \cite{gerry2002hole}. Interestingly, this result also implies that
qudits which are $d$-dimensional (finite dimensional) quantum states
are always nonclassical as we can see that in  such a state $p_{d}=p_{d+1}=\ldots=0$. In principle, the hole can be created for an arbitrary $n$, but here
for the sake of a comparative study, we restrict ourselves to the situation
where the hole is created at $n=0,$ i.e., the desired engineered state has zero probability of getting vacuum state on measurement
in Fock basis (in other words, $p_{0}=0).$ In fact,
Lee \cite{lee1995theorem} had shown that a state with $p_{n}=0$
is a maximally nonclassical as long as the nonclassicality is quantitatively
measured using nonclassical depth. Such a state can be constructed in
various ways. To elaborate on this, we describe an arbitrary pure quantum state as a  superposition
of the Fock states 
\begin{equation}
|\psi\rangle=\sum\limits _{n=0}^{\infty}c_{n}|n\rangle,\label{eq:fock-superposition}
\end{equation}
where $c_{n}$ is the probability amplitude of state $|n\rangle$.
A hole can be created at $n=0$ by adding a single photon  to obtain 
\begin{equation}
|\psi_{1}\rangle=N_1 a^{\dagger}|\psi\rangle,\label{eq:a-dagger-for-hole}
\end{equation}
where $N_1=\left(\langle \psi_{1}|aa^{\dagger} |\psi_{1}\rangle\right)^{-1/2}$ is the normalization constant.
If we consider the initial quantum state $|\psi\rangle$ as a coherent
state, the addition of a single photon would lead to a
photon added coherent state which has been experimentally realized
\cite{zavatta2004quantum} and extensively studied
\cite{hong1999nonclassical} because of its interesting
nonclassical properties and potential applications. Thus, in quantum state engineering, techniques
for photon addition are known \cite{thapliyal2017comparison,malpani2019lower,malpani2019quantum} and  experimentally realized.

An alternative technique to create a hole at vacuum is vacuum filtration.
The detailed procedure of this technique is recently discussed in
\cite{Meher2018}. Vacuum filtration implies removal of the coefficient
of the vacuum state, $c_{0}$, in Eq. (\ref{eq:fock-superposition}) and
subsequent normalization. Clearly this procedure would yield 
\begin{equation}
|\psi_{2}\rangle=N_2\sum\limits _{n=1}^{\infty}c_{n}^{\prime}|n\rangle,\label{eq:c0-is-removed}
\end{equation}
where the normalization
constant $N_2=\left(1-\left|c_{0}\right|^{2}\right)^{-1/2}$. 
Both these states (i.e., $|\psi_{1}\rangle,$ and $\psi_{2}\rangle$) are maximally nonclassical as far as Lee's
result related to nonclassical depth is concerned \cite{lee1991measure}.
However, recently lower-order nonclassical
properties of $|\psi_{1}\rangle$ and $|\psi_{2}\rangle$ in (\ref{eq:a-dagger-for-hole})-(\ref{eq:c0-is-removed})
are reported to be different for $|\psi\rangle$ chosen as coherent
state \cite{Meher2018}.    This led to several interesting questions, like- What happens if the initial state
on which addition of photon or vacuum filtration process is to be
applied is already nonclassical (specifically, pure state other than coherent state \cite{hillery1985classical})? How do these processes affect the
higher-order nonclassical properties of the quantum states? How does
the depth of nonclassicality corresponding to a particular witness
of nonclassicality changes with the parameters of the quantum state
for these processes?
 The present article
aims to answer these questions through a comparative study using a
set of interesting quantum states $|\psi\rangle$ (and the corresponding single photon added $|\psi_{1}\rangle$ and vacuum filtered $|\psi_{2}\rangle$ states), each of which can be reduced to many more states.
Specifically, in what follows, we would study the lower- and higher-order
nonclassical properties of single photon addition and vacuum filtration
of even coherent state (ECS), binomial state (BS) and Kerr state (KS). In fact, the quantum state engineering
processes described mathematically in Eqs. (\ref{eq:a-dagger-for-hole})-(\ref{eq:c0-is-removed})
can be used to prepare a set of engineered quantum states, namely vacuum filtered
ECS (VFECS), vacuum filtered BS (VFBS), vacuum filtered KS (VFKS),
photon added ECS (PAECS), photon added BS (PABS) and photon added
(PAKS). We aim to look at the nonclassical properties
of these states with a focus on higher-order nonclassical properties and subsequently quantify the amount of nonclassicality in all these states.
In what follows, the higher-order
nonclassical properties are illustrated through the criteria of higher-order antibunching (HOA), higher-order squeezing
(HOS) and higher-order sub-Poissonian photon statistics (HOSPS) with brief discussion of lower-order antibunching and squeezing. 

The rest of the paper is organized as follows. In Section \ref{sec:Quantum-states-of},
we have introduced the quantum states of our interest which include
ECS, BS, KS, VFECS, VFBS, VFKS, PAECS, PABS, and PAKS. In Section \ref{sec:Nonclassicality-witnesses},
we have investigated the nonclassical properties of these states using
various witnesses of lower- and higher-order noncassicality as well
as a measure of nonclassicality. Specifically, in this section, we
have compared nonclassicality features found in vacuum filtered and
single photon added versions of the states of our interest using the witnesses
of HOA, HOS and HOSPS. Finally,
in Section \ref{sec:Conclusion}, the results are analyzed, and the
paper is concluded. 

\section{Quantum states of interest\label{sec:Quantum-states-of}}

In this paper, we have selected a set of three widely studied and
important quantum states- (i) ECS, (ii) BS
and (iii) KS. We subsequently noted that these states
can further be engineered to generate corresponding vacuum filtered states
and single photon added states. For example, one can
generate VFBS and PABS from BS by using vacuum filtration \cite{Meher2018}
and photon addition \cite{zavatta2004quantum} processes, respectively.
In a similar manner, these processes can also generate VFECS and PAECS
from ECS, and VFKS and PAKS from KS. In this section, we briefly describe ECS, BS, KS, 
VFBS, PABS, VFECS, PAECS, VFKS and PAKS. Specifically, we describe three parent states as Fock superposition states. Similarly,
the six engineered states are also expressed as Fock superposition states for the convenience
of identifying the corresponding photon number distributions (each
of which essentially contains a hole at the vacuum). In the rest of the
study, we wish to compare the impact of these two quantum state
engineering processes (i.e., vacuum filtration and photon
addition processes) on the nonclassical properties of the engineered states. 

\subsection{Even coherent state and states generated by holeburning on it}
The analytical expression for ECS in number basis
can be written as 
\begin{equation}
\begin{array}{lcl}
|\phi(\alpha)\rangle & = & \frac{\,\exp\left[-\frac{\mid\alpha\mid^{2}}{2}\right]}{\sqrt{2\left(1+\exp\left[-2\mid\alpha\mid^{2}\right]\right)}}\sum\limits _{n=0}^{\infty}\frac{\alpha^{n}}{\sqrt{n!}}\left(1+\left(-1\right)^{n}\right)|n\rangle.\end{array}\label{eq:ECS}
\end{equation}
The parameter $\alpha=|\alpha|\exp (i\theta)$, in Eqs. (\ref{eq:ECS}),
is complex in general and $\theta$ is phase angle in the complex plane.

Various schemes to generate ECS are reported in \cite{brune1992manipulation,ourjoumtsev2007generation,gerry1993non}.
The nonclassical properties (witnessed through the antibunching and squeezing criteria, $Q$ function, Wigner
function, and photon number distribution, etc.) of ECS have been studied
in the recent past  \cite{gerry1993non}. In
what follows, we study, both qualitatively and quantitatively, the role of vacuum filtration and photon addition on the nonclassical
properties of VFECS and PAECS and compare them with the corresponding properties of  ECS. 

\subsubsection{Vacuum filtered even coherent state}

Experimentally, an ECS or a cat state can be generated in various
ways \cite{ourjoumtsev2007generation}, and the same
can be further engineered to produce a hole at vacuum in its photon
number distribution. Specifically, filtration of vacuum will burn
a hole at $n=0$ and produce VFECS, which can described in Fock basis
as
\begin{equation}
\begin{array}{lcl}
|\phi_{1}(\alpha)\rangle & = & N_{{\rm VFECS}}\sum\limits _{n=0,\,n\neq0}^{\infty}\frac{\alpha^{n}}{\sqrt{n!}}\left(1+\left(-1\right)^{n}\right)|n\rangle,\end{array}\label{eq:VFECS}
\end{equation}
where
\begin{equation}
\begin{array}{lcl}
N_{{\rm VFECS}} & = & \{4{\rm cosh}\left(\mid\alpha\mid^{2}\right)-1\}^{-1/2}\end{array}
\end{equation}
is the normalization constant. For simplicity, we may expand Eq. (\ref{eq:VFECS})
as a superposition of a standard ECS and a vacuum state as follows
\begin{equation}
\begin{array}{lcl}
|\phi_{1}(\alpha)\rangle & = & N_{{\rm VFECS}}\left(\sum\limits _{n=0}^{\infty}\frac{\alpha^{n}}{\sqrt{n!}}\left(1+\left(-1\right)^{n}\right)|n\rangle-2|0\rangle\right).\end{array}\label{eq:VFECS-EXPANDED}
\end{equation}
In what follows, Eq. (\ref{eq:VFECS-EXPANDED}) will be used to explore
various nonclassical features that exist in VFECS. Specifically, we
will compute a general expression for moment of annihilation and creation
operators for this state and use a set of moment-based criteria of
nonclassicality to identify the nonclassical properties of this engineered
quantum state. Similar prescription will be followed for the other
engineered quantum states of our interest.

\subsubsection{Photon added even coherent state}

One can define a single photon added ECS as
\begin{equation}
|\phi_{2}(\alpha)\rangle=N_{{\rm PAECS}}\hat{a}^{\dagger}|\phi(\alpha)\rangle=N_{{\rm PAECS}}\sum\limits _{n=0}^{\infty}\frac{\alpha^{n}}{\sqrt{n!}}\left(1+\left(-1\right)^{n}\right)\sqrt{n+1}|n+1\rangle,\label{eq:PAECS}
\end{equation}
where
\begin{equation}
\begin{array}{lcl}
N_{{\rm PAECS}} & = & \{{\rm cosh}\left(\mid\alpha\mid^{2}\right)+\mid\alpha\mid^{2}{\rm sinh}\left(\mid\alpha\mid^{2}\right)\}^{-1/2}/2\end{array}
\end{equation}
 is the normalization constant for PAECS. 

\subsection{Binomial state   and the states generated by holeburning on it}

BS is a finite superposition of Fock states having
binomial photon number distribution. It is quite similar to the coherent
state which is the linear combination of Fock states having the Poissonian
photon number distribution \cite{stoler1985binomial}. BS can be defined as
\begin{equation}
\begin{array}{lcl}
|p,M\rangle & = &\sum\limits _{n=0}^{M}\left[\frac{M!}{n!(M-n)!}p^{n}\left(1-p\right)^{M-n}\right]^{1/2}|n\rangle.\end{array}\label{eq:BS}
\end{equation}
 The binomial coefficient describes the presence of $n$ photons with probability $p$ in $M$ number of ways. Recently, some of the present authors
have extensively studied the nonclassical properties of BS, specifically, antibunching, squeezing, HOSPS \cite{verma2008higher,verma2010generalized,bazrafkan2004tomography},
etc. However, no effort has yet been made to study the nonclassical properties of VFBS and PABS. 

\subsubsection{Vacuum filtered Binomial state}

The vacuum filtration of BS can be obtained by simply
eliminating vacuum state from the BS as
\begin{equation}
\begin{array}{lcl}
|p,M\rangle_1 & = & N_{{\rm VFBS}}\sum\limits _{n=0}^{M}\left[\frac{M!}{n!(M-n)!}p^{n}\left(1-p\right)^{M-n}\right]^{1/2}|n\rangle-N_{VFBS}\left[\left(1-p\right)^{M}\right]^{1/2}|0\rangle,\end{array}\label{eq:VFBS}
\end{equation}
where
\begin{equation}
\begin{array}{lcl}
N_{{\rm VFBS}} & = & \{1-\left(1-p\right)^{M}\}^{-1/2}\end{array}
\end{equation}
is the normalization constant for the VFBS. 

\subsubsection{Photon added Binomial state}

A hole at $n=0$ at a BS can also be introduced by the addition of a single
photon on the BS. A few steps of computation yield the
desired expression for PABS as
\begin{equation}
\begin{array}{lcl}
|p,M\rangle_2 & =N_{{\rm PABS}} & \sum\limits _{n=0}^{M}\left[\frac{M!(n+1)!}{\left(n!\right)^{2}(M-n)!}p^{n}\left(1-p\right)^{M-n}\right]^{1/2}|n+1\rangle,\end{array}\label{eq:PABS}
\end{equation}
where
\begin{equation}
\begin{array}{lcl}
N_{{\rm PABS}} & = & \left(1+Mp\right)^{-1/2}\end{array}
\end{equation}
is the normalization constant for single photon added BS. 

\subsection{Kerr state  and the states generated by holeburning on it}

A KS can be obtained when  a coherent state of electromagnetic
field interacts with nonlinear medium with Kerr type nonlinearity \cite{gerry1994statistical}. This interaction generates phase shifts
which depend on the intensity. The Hamiltonian involved in this process
is given as 
\begin{equation}
H=\hbar\omega\hat{a}^{\dagger}\hat{a}+\hbar\chi\left(\hat{a}^{\dagger}\right)^{2}\left(\hat{a}\right)^{2},
\end{equation}
where $\chi$ depends on the third-order susceptibility of Kerr medium. Thus, the compact
analytic form of the KS in the Fock basis can be given as
\begin{equation}
\begin{array}{lcl}
|\psi_{K}\left(n\right)\rangle & = & \sum\limits _{n=0}^{\infty}\frac{\alpha^{n}}{\sqrt{n!}}\exp\left(-\frac{\mid\alpha\mid^{2}}{2}\right)\exp\left(-\iota\chi n\left(n-1\right)\right)|n\rangle.\end{array}\label{eq:KS}
\end{equation}

\subsubsection{Vacuum filtered Kerr state}

Similarly, a VFKS, which can be obtained using the same quantum state engineering
process that leads to VFECS and VFBS, is given by
\begin{equation}
\begin{array}{lcl}
|\psi_{K}\left(n\right)\rangle_{1} & = & N_{{\rm VFKS}}\left[\sum\limits _{n=0}^{\infty}\frac{\alpha^{n}}{\sqrt{n!}}\exp\left(-\iota\chi n\left(n-1\right)\right)|n\rangle-|0\rangle\right]\end{array},\label{eq:VFKS-expanded}
\end{equation}
where
\begin{equation}
\begin{array}{lcl}
N_{{\rm VFKS}} & = &{{\left(\exp\left[\mid\alpha\mid^{2}\right]-1\right)}^{-1/2}}\end{array}
\label{eq:NVFKS-expanded}
\end{equation}
is the normalization constant for the VFKS. 

\subsubsection{Photon added Kerr state}

An addition of a photon to KS would yield PAKS which can be
expanded in Fock basis as 
\begin{equation}
\begin{array}{lcl}
|\psi_{K}\left(n\right)\rangle_{2} & =N_{{\rm PAKS}} & \sum\limits _{n=0}^{\infty}\frac{\alpha^{n}}{\sqrt{n!}}\exp\left(-\iota\chi n\left(n-1\right)\right)\sqrt{\left(n+1\right)}|n+1\rangle,\end{array}\label{eq:PAKS}
\end{equation}
where
\begin{equation}
\begin{array}{lcl}
N_{{\rm PAKS}} & = & {\left(\exp\left[\mid\alpha\mid^{2}\right]\left(1+\mid\alpha\mid^{2}\right)\right)^{-1/2}}\end{array}
\label{eq:NPAKS}
\end{equation}
is the normalization constant for the PAKS. 

In the above, we have described six (three) quantum states of our interest
as Fock superposition states having (without) holes at vacuum. In what follows, these expressions will
be used to study the nonclassical properties of these states using
a set of witnesses of nonclassicality. Specifically, we will use a
set of witnesses of nonclassicality which are based on moments of
annihilation and creation operators. Keeping this in mind, in the following subsection, we report
the general form of such moments for all the six engineered states of our
interest and the corresponding three parent states (thus overall nine states).

\subsection{Expressions for moments of annihilation and creation operators}

In 1992, Agarwal and Tara \cite{agarwal1992nonclassical} introduced
a criterion of nonclassicality in the form of a matrix of moments of creation
and annihilation operators. This criterion was further modified 
to propose a moment-based criteria of entanglement
\cite{shchukin2005inseparability} and nonclassicality \cite{miranowicz2010testing,miranowicz2009inseparability}.
Therefore, it is convenient to find out the expectation value
of the most general term describing higher-order moment $\langle\hat{a}^{\dagger j}\hat{a}^{k}\rangle$ for a given state to investigate the nonclassicality using the set
of moment-based criteria. 

\subsubsection{Expectation values for even coherent
states  and the corresponding engineered states}

The analytic expression of $\langle\hat{a}^{\dagger j}\hat{a}^{k}\rangle_{i}$
is obtained for the quantum states $i\in\{{\rm ECS},{\rm VFECS,PAECS\}}$
using Eqs. (\ref{eq:VFECS-EXPANDED}) and (\ref{eq:PAECS}). For ECS and VFECS,
expressions of the moments can be written in a compact form as 

\begin{equation}
\begin{array}{cc}
\langle\hat{a}^{\dagger j}\hat{a}^{k}\rangle_{{\rm ECS}}= & \frac{\exp\left[-\mid\alpha\mid^{2}\right]}{2\left(1+\exp\left[-2\mid\alpha\mid^{2}\right]\right)}\sum\limits _{n=0}^{\infty}\frac{\alpha^{n}(\alpha^{\star})^{n-k+j}}{\left(n-k\right)!}\left(1+\left(-1\right)^{n}\right)\left(1+\left(-1\right)^{n-k+j}\right).\end{array}\label{eq:ECS-moment}
\end{equation}
and
\begin{equation}
\begin{array}{ccc}
\langle\hat{a}^{\dagger j}\hat{a}^{k}\rangle_{{\rm VFECS}} & = & \left\{ \begin{array}{c}
N_{{\rm VFECS}}^{2}\sum\limits _{n=1}^{\infty}\frac{\alpha^{n}(\alpha^{\star})^{n-k+j}}{\left(n-k\right)!}\left(1+\left(-1\right)^{n}\right)\left(1+\left(-1\right)^{n-k+j}\right)\,\,\,\,\mathrm{for}\,\,k\leq j,\\
N_{{\rm VFECS}}^{2}\sum\limits _{n=1}^{\infty}\frac{\alpha^{\star n}\alpha^{n+k-j}}{\left(n-j\right)!}\left(1+\left(-1\right)^{n}\right)\left(1+\left(-1\right)^{n+k-j}\right)\,\,\,\mathrm{for}\,\,k>j,
\end{array}\right.\end{array}\label{eq:VFECS-moment}
\end{equation}
respectively. 
Similarly, we obtained analytic expression for $\langle\hat{a}^{\dagger j}\hat{a}^{k}\rangle_{{\rm PAECS}}$
for PAECS as 
\begin{equation}
\begin{array}{cl}
\langle\hat{a}^{\dagger j}\hat{a}^{k}\rangle_{{\rm PAECS}}= & N_{{\rm PAECS}}^{2}\sum\limits _{n=0}^{\infty}\frac{\alpha^{n}(\alpha^{\star})^{n-k+j}\left(n+1\right)\left(n-k+j+1\right)}{\left(n+1-k\right)!}\left(1+\left(-1\right)^{n}\right)\left(1+\left(-1\right)^{n-k+j}\right).\end{array}\label{eq:PAECS-moment}
\end{equation}
 The above mentioned quantities are also functions of displacement
parameter of ECS used to generate the engineered states, which will be
used as a control parameter while discussion of nonclassicality induced
due to engineering operations.

\subsubsection{Expectation values for binomial state and the corresponding engineered states}

Similarly, the compact analytic form of $\langle\hat{a}^{\dagger t}\hat{a}^{r}\rangle_{{\rm BS}}$
can be written as
\begin{equation}
\begin{array}{lcl}
\langle\hat{a}^{\dagger t}\hat{a}^{r}\rangle_{{\rm BS}} & = & \sum\limits _{n=0}^{M}\left[\frac{p^{2n-r+t}\left(1-p\right)^{2M-2n+r-t}}{(M-n)!(M-n+r-t)!}\right]^{1/2}\frac{M!}{(n-r)!}.\end{array}\label{eq:BS-moment}
\end{equation}
In case of VFBS and PABS, the analytic form of $\langle\hat{a}^{\dagger t}\hat{a}^{r}\rangle_{{i}}$
is obtained as 
\begin{equation}
\begin{array}{lcl}
\langle\hat{a}^{\dagger t}\hat{a}^{r}\rangle_{{\rm VFBS}} & = &\left\{ \begin{array}{c} N_{{\rm VFBS}}^{2}\sum\limits _{n=1}^{M}\left[\frac{p^{2n-r+t}\left(1-p\right)^{2M-2n+r-t}}{(M-n)!(M-n+r-t)!}\right]^{1/2}\frac{M!}{(n-r)!}\,\,\,\,\,\rm{for}\,r\leq t,\\
 N_{{\rm VFBS}}^{2}\sum\limits _{n=1}^{M}\left[\frac{p^{2n+r-t}\left(1-p\right)^{2M-2n-r+t}}{(M-n)!(M-n-r+t)!}\right]^{1/2}\frac{M!}{(n-t)!}\,\,\,\,\,\rm{for}\,r>t, \end{array} \right.
\end{array}\label{eq:VFBS-moment}
\end{equation}
and 
\begin{equation}
\begin{array}{lcl}
\langle\hat{a}^{\dagger t}\hat{a}^{r}\rangle_{{\rm PABS}} & = & N_{{\rm PABS}}^{2}\sum\limits _{n=0}^{M}\left[\frac{p^{2n-r+t}\left(1-p\right)^{2M-2n+r-t}}{(M-n)!(M-n+r-t)!}\right]^{1/2}\frac{M!(n+1)!(n+1-r+t)!}{n!(n+1-r)!(n-r+t)!},\end{array}\label{eq:PABS-moment}
\end{equation}
respectively. Here, the obtained average values of moments are also dependent on
BS parameters, which will be used to enhance/control the nonclassicality
features in the generated states.

\subsubsection{Expectation values for Kerr state  and the corresponding engineered states}

For KS, VFKS and PAKS, we use the same approach to obtain a compact generalized
forms of $\langle\hat{a}^{\dagger q}\hat{a}^{s}\rangle_{{i}}$; and our computation yielded
\begin{equation}
\begin{array}{lcl}
\langle\hat{a}^{\dagger q}\hat{a}^{s}\rangle_{{\rm KS}} & = & \sum\limits _{n=0}^{\infty}\frac{\alpha^{n}(\alpha^{\star})^{n-s+q}}{\left(n-s\right)!}\exp\left[-\mid\alpha\mid^{2}\right]\exp\left(\iota\chi\left[\left(n-s+q\right)\left(n-s+q-1\right)-n\left(n-1\right)\right]\right),
\end{array}\label{eq:kS-moment}
\end{equation}
\begin{equation}
\begin{array}{l}
\langle\hat{a}^{\dagger q}\hat{a}^{s}\rangle_{{\rm VFKS}}=\left\{ \begin{array}{c}
N_{{\rm VFKS}}^{2}\sum\limits _{n=1}^{\infty}\frac{\alpha^{n}(\alpha^{\star})^{n-s+q}}{\left(n-s\right)!}\exp\left(\iota\chi\left[\left(n-s+q\right)\left(n-s+q-1\right)-n\left(n-1\right)\right]\right),\,{\rm {for}\,\,s\leq q,}\\
N_{{\rm VFKS}}^{2}\sum\limits _{n=1}^{\infty}\frac{\alpha^{\star n}\alpha^{n+s-q}}{\left(n-q\right)!}\exp\left(-\iota\chi\left[\left(n+s-q\right)\left(n+s-q-1\right)-n\left(n-1\right)\right]\right),\,{\rm {for}\,\,s>q,}
\end{array}\right.\end{array}\label{eq:VFkS-moment}
\end{equation}
and 
\begin{equation}
\begin{array}{lcl}
\langle\hat{a}^{\dagger q}\hat{a}^{s}\rangle_{{\rm PAKS}} & = & N_{{\rm PAKS}}^{2}\sum\limits _{n=0}^{\infty}\frac{\alpha^{n}(\alpha^{\star})^{n-s+q}\left(n+1\right)!\left(n-s+q+1\right)!}{n!\left(n-s+q\right)!(n+1-s)!} \exp\left(\iota\chi\left[\left(n-s+q\right)\left(n-s+q-1\right)-n\left(n-1\right)\right]\right).
\end{array}\label{eq:PAKS-moment}
\end{equation}
From the above expressions, it is clear that when $q=s$, there is
no role of $\chi$ and the behavior of KS is similar to that of a coherent
state. So the effect of this parameter $\left(\chi\right)$ can be
observed only in HOS which also depends on the higher-order moments other than moments of number operator, i.e., $\langle\hat{a}^{\dagger q}\hat{a}^{s}\rangle_{{i}} \,: q\neq s$. In what follows,
we use the expressions of moments given in Eqs. (\ref{eq:VFECS-moment})-(\ref{eq:PAKS-moment})
to study various lower- and higher-order nonclassicality
witnesses. 

\section{Nonclassicality witnesses\label{sec:Nonclassicality-witnesses}}

There are various criteria of nonclassicality, most of them are sufficient
but not necessary in the sense that satisfaction of such a criterion
can identify a nonclassical feature, but failure does not ensure that
the state is classical. Further, most of the criteria (specially, all
the criteria studied here) do not provide any quantitative measure of nonclassicality present in a state, and so they are referred to
as witnesses of nonclassicality. These witnesses are based on either
quasiprobability distribution or moments of annihilation and creation
operators. In the present work, we have used a set of moment-based
criteria to investigate nonclassical properties of our desired engineered quantum states. Specifically, we have investigated
the possibilities of observing lower-order squeezing and antibunching as well as HOA,
HOSPS, and HOS
for all the states of our interest. To begin the investigation
and the comparison process, let us start with the study of antibunching.

\subsection{Lower- and higher-order antibunching}

\begin{figure}
\centering{}
\includegraphics[scale=0.7]{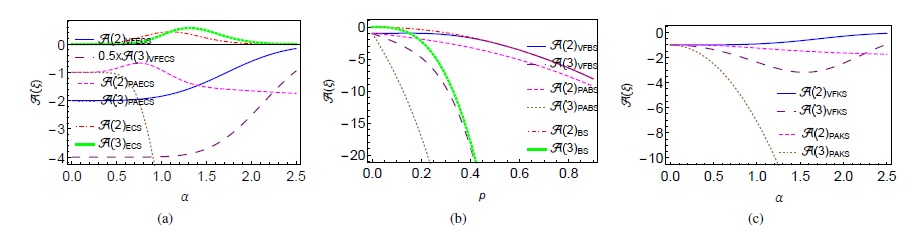}
\caption{\label{fig:HOA}(Color online) Lower- and higher-order antibunching witnesses
as functions of displacement parameter $\alpha$ (for ECS and KS)
and probability $p$ (for BS with parameter $M=10$) for (a) ECS,
PAECS and VFECS, (b) BS, PABS and VFBS, and (c) KS, PAKS and VFKS. The quantities shown in all the plots are dimensionless.}
\end{figure}

The phenomenon of lower-order antibunching is closely associated with
the lower-order sub-Poissonian photon statistics
\cite{brown1956correlation}. However, they are not equivalent \cite{zou1990photon}. The concept of HOA (higher-order nonclassicality)
also plays an important role in identifying the presence
of weaker nonclassicality \cite{allevi2012measuring,hamar2014non}.
It was first introduced in 1990 based on majorization technique
\cite{lee1990higher} followed by some of its modifications  
\cite{an2002multimode,pathak2006control}.
In this section, we study the generalized HOA criterion introduced
by Pathak and Garcia \cite{pathak2006control} to investigate lower-order antibunching
and HOA. To do so, we use the following criterion
for $\xi^{{\rm th}}$ order antibunching \cite{pathak2006control,gupta2006higher}
\begin{equation}
\mathscr{A}(\xi)=\langle\hat{a}^{\dagger\left(\xi+1\right)}\hat{a}^{\left(\xi+1\right)}\rangle-\langle\hat{a}^{\dagger}\hat{a}\rangle^{\xi+1}<0,\label{eq:HOA}
\end{equation}
where $\xi$ is a positive integer. Depending upon the values of $\xi$,
Eq. (\ref{eq:HOA}) reduces to lower- and higher-order criteria of
antibunching for $\xi=1$ and $\xi\geq2$,
respectively. The analytic expressions of moments (\ref{eq:VFECS-moment})-(\ref{eq:PAKS-moment}) can be used
to investigate the nonclassicality using inequality (\ref{eq:HOA}) for the set of states.
The obtained results are illustrated in Fig. \ref{fig:HOA} where
we have compared the results between the vacuum filtered and single photon
added states. During this attempt, we also discuss the nonclassicality present
in the quantum states used for the preparation of the engineered quantum states (cf. Figs. \ref{fig:HOA} (a)-(c)).
In Figs. \ref{fig:HOA} (b)-(c), we have shown the result for photon
added and vacuum filtered BS and KS, where it can be observed that
the depths of both lower- and higher-order witnesses in the
negative region are larger for photon added BS and KS in comparison
with the vacuum filtered BS and KS, respectively. However, an opposite nature is observed
for ECS where the depth of lower- and higher-order
witnesses is more for the vacuum filtration in comparison with the
photon addition if the values of $\alpha$ remain below certain values;
whereas for the photon addition the depth of lower- and higher-order
antibunching witnesses is found to be greater than that for vacuum filtration
for the higher values of $\alpha$ (cf. Fig. \ref{fig:HOA} (a)).
However, HOA is not observed for the ECS and KS and thus both operations can be ascribed as nonclassicality inducing operations as far as this nonclassical feature is concerned. 

%-----------------    Figure-1   -----------------------------------
%--------------------------------------------------------------------------------

\subsection{Lower- and higher-order squeezing}

The concept of squeezing originates from the uncertainty relation.
There is a minimum value of an uncertainty relation involving quadrature
operators where the variance of two non-commuting quadratures (say
position and momentum) are equal and their product satisfies minimum
uncertainty relation. Such a situation is closest to the classical scenario,
in the sense that there is no uncertainty in the classical picture
and this is the closest point that one can approach remaining within
the framework of quantum mechanics. Coherent state satisfies this
minimum uncertainty relation and is referred to as a classical (or
more precisely closest to classical state). If any of the quadrature
variances reduces below the corresponding value for a minimum uncertainty
(coherent) state (at the cost of increase in the fluctuations in the
other quadrature) then the corresponding state is called squeezed
state. 

\begin{figure}
\centering{}
\includegraphics[scale=0.7]{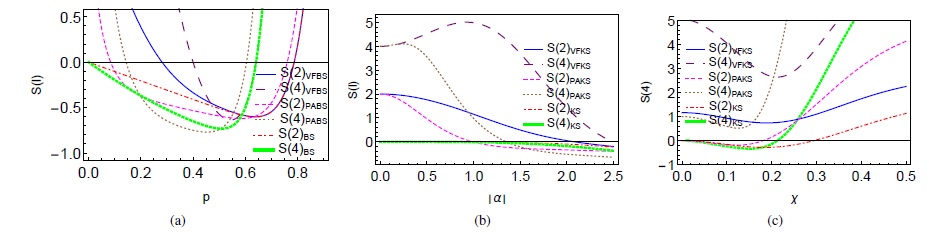}
\caption{\label{fig:HOS}(Color online) Illustration of lower- and higher-order squeezing for (a) BS, PABS
and VFBS; (b) KS, PAKS and VFKS at the fixed value of $\chi=0.02$; (c) KS, VFKS and PAKS as a function of $\chi$ with
$\alpha=1$. The negative regions of the curves illustrate   the presence
of squeezing.}  
\end{figure}

The higher-order nonclassical properties can be investigated
by studying HOS. There are two different criteria for HOS \cite{hong1985generation,hong1985higher,hillery1987amplitude}:
Hong-Mandel criterion \cite{hong1985higher} and Hillery criterion
\cite{hillery1987amplitude}. The concept of the HOS was first introduced
by Hong and Mandel using higher-order moments of the quadrature operators
\cite{hong1985higher}. According to this criterion, it is observed
if the higher-order moment for a quadrature operator for   a quantum
state is observed to be less than the corresponding coherent state
value. Another   type of HOS was introduced by Hillery who introduced
amplitude powered quadrature and used variance of this quadrature
to define HOS \cite{hillery1987amplitude}. Here, we aim to analyze
the possibility of HOS using Hong-Mandel criterion for $l$th order
squeezing defined as \cite{hong1985higher}
\begin{equation}
S(l)=\frac{\langle\left(\text{\ensuremath{\Delta}}\text{X}\right)^{l}\rangle-\left(\frac{1}{2}\right)_{\frac{l}{2}}}{\left(\frac{1}{2}\right)_{\frac{l}{2}}}<0,\label{eq:HOS}
\end{equation}
where $X=\left(a+a^{\dagger}\right)/\sqrt{2}$ is the dimensionless
quadrature, the  symbol $\left(x\right)_{n}$ is the conventional Pochhammer
symbol, and $l$ is an even integer >2 for HOS, which represents the order of the
squeezing.  Computation
of $\langle\left(\text{\ensuremath{\Delta}}\text{X}\right)^{l}\rangle$
was a bit tedious and HOS of Hong-Mandel type was reported only in
a few states until the recent past when Verma and Pathak
\cite{verma2010generalized} reduced the complexity associated with
the operator ordering involved in the computation of $\langle\left(\text{\ensuremath{\Delta}}\text{X}\right)^{l}\rangle$
and provided the following $c$-number-based criterion for Hong-Mandel type HOS which
can be computed easily from the expressions of moments provided in
the previous section
\begin{equation}
\begin{array}{ccccc}
\langle\left(\text{\ensuremath{\Delta}}\text{X}\right)^{l}\rangle & = & \sum\limits _{r=0}^{l}\sum\limits _{i=0}^{\frac{r}{2}}\sum\limits _{k=0}^{r-2i}\left(-1\right)^{r}\frac{1}{2^{\frac{l}{2}}}\left(2i-1\right)!^{2i}C_{k}{}^{l}C_{r}{}^{r}C_{2i}\langle\hat{a}^{\dagger}+\hat{a}\rangle^{l-r}\langle\hat{a}^{\dagger k}\hat{a}^{r-2i-k}\rangle<\left(\frac{1}{2}\right)_{\frac{l}{2}} & = & \frac{1}{2^{\frac{l}{2}}}\left(l-1\right)!!.\end{array}\label{eq:HOS-1}
\end{equation}
Note that $l=2$ corresponds to lower-order squeezing. We have investigated the possibility of observing HOS analytically
using Eqs. (\ref{eq:VFECS-moment})-(\ref{eq:PAKS-moment}) and inequality
(\ref{eq:HOS-1}) for all engineered quantum states of our
interest and have shown the corresponding results in Figs. \ref{fig:HOS} (a)-(c)
where we have compared the HOS in the set of quantum states and the
states obtained by  photon addition and vacuum filtration. These operations fail to induce this nonclassical feature in the engineered states prepared from ECS, which also did not show signatures of squeezing.
In Fig. \ref{fig:HOS} (a), we illustrate Hong-Mandel type HOS with
respect to parameter $p$ where we have shown the existence HOS
for BS, VFBS and PABS. It can be observed that the state engineering operations fail to increase this particular feature of nonclassicality in BS. Additionally, higher-order nonclassicality is absent for higher values of $p$ when corresponding lower-order squeezing is present. In case of KS, PAKS and VFKS, we have observed
that HOS is observed when the values of $\alpha$ are greater than
certain values for the individual curves of the corresponding states
(cf. Fig. \ref{fig:HOS} (b)). Note that photon addition may provide some advantage in this case,  but vacuum filtration would not as for the same value of displacement parameter KS and PAKS (VFKS) have (has not) shown squeezing. Interestingly, the presence of squeezing also depends upon the Kerr nonlinearity parameter $\chi$, which is shown in Fig. \ref{fig:HOS} (c). Similar to Fig. \ref{fig:HOS} (b) photon addition shows advantage over KS which disappears for larger values of $\chi$, while vacuum filtering is not beneficial.

\begin{figure}
\centering{}
\includegraphics[scale=0.9]{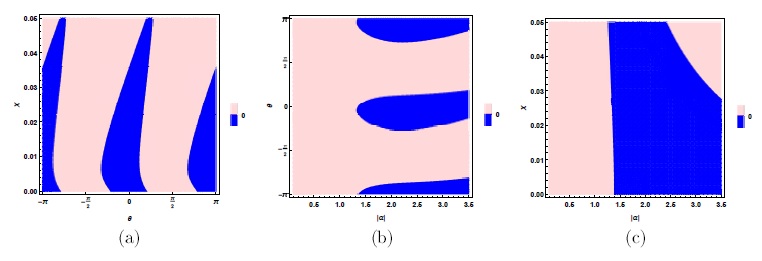}
\caption{\label{fig:HOS-contour}(Color online) The dependence of HOS witness ($l=4$) on Kerr parameter $\chi$ and displacement
parameters $\left|\alpha\right|$ and $\theta$ for PAKS with (a) $\left|\alpha\right|=3$, (b) $\chi=0.02$,
(c) $\theta=0$.}
\end{figure}

In Fig. \ref{fig:HOS-contour},
we have shown using the dark (blue) color in the contour plots of the HOS witness for PAKS that squeezing can be observed for higher values of $|\alpha|$ and smaller values of $\chi$. Additionally, the phase parameter $\theta$ of $\alpha$ is also relevant for observing the nonclassicality as squeezing occurs in the vicinity of $\theta=m\pi$, while disappears for $\theta=\frac{m\pi}{2}$ with integer $m$. Similar behavior is observed in KS and VFKS (not shown here). 

%--------------- Figure-2 --------------------------------------------------
%---------------------------------------------------------------------------------

%---------------------  Figure-3  -----------------------------------------------
%---------------------------------------------------------------------------------

\subsection{Higher-order sub-Poissonian photon statistics}

\begin{figure}
\centering{}
\includegraphics[scale=0.7]{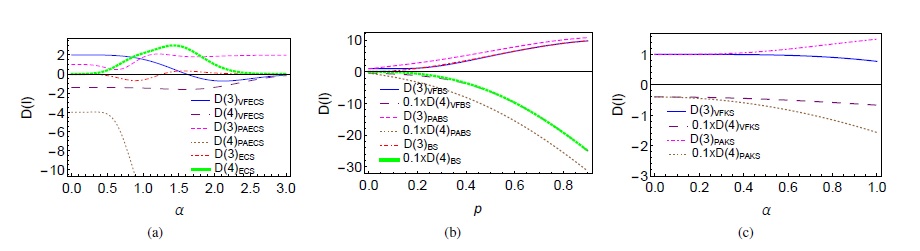}
\caption{\label{fig:HOSPS}(Color online) Illustration of HOSPS as a function
of displacement parameter $\alpha$ (for ECS and KS) and probability
$p$ (for BS) for (a) ECS, (b) BS, and (c) KS and corresponding engineered states. HOSPS is not observed in KS.}
\end{figure}

Quantum statistical properties of a radiation field can be investigated
through HOSPS. Mathematically,
it is observed if the higher-order variance of the photon number is
less than its Poissonian level ($\langle\left(\Delta N\right)^{l}\rangle<\langle\left(\Delta N\right)^{l}\rangle_{{\rm Poissonian}}$).
Moreover, HOSPS is found useful in various aspects of higher-order
nonclassical phenomena \cite{prakash2006higher}.
The condition for HOSPS is defined as 
\begin{equation}
\begin{array}{lcccc}
{D}(l) & = & \sum\limits _{u=0}^{l}\sum\limits _{v=0}^{u}S_{2}(u,\,v)\,^{l}C_{u}\,\left(-1\right)^{u}\mathscr{A}(v)\langle N\rangle^{l-u} & < & 0.\end{array}\label{eq:hosps22}
\end{equation}
where $S_{2}(u,\,v)$ stands for the Stirling numbers of second kind,
$N$ is the usual number operator. The inequality in Eq. $(\ref{eq:hosps22})$
is the condition for the $l$th order sub-Poissonian photon statistics
representing corresponding higher-order counterparts for $l>2$.
The higher-order moments in Eqs. (\ref{eq:VFECS-moment})-(\ref{eq:PAKS-moment})
are used to calculate the above inequality $(\ref{eq:hosps22})$ with the help of  (\ref{eq:HOA}) for states obtained after
vacuum filtration and photon addition in ECS, BS and KS as well
as the parent states, and the corresponding results are depicted in
Fig. \ref{fig:HOSPS}. Nonclassicality is not revealed by HOSPS criteria of even orders in case of ECS, while corresponding engineered states show nonclassicality. Additionally, nonclassicality is induced by vacuum filtration for odd orders while it was not observed in the parent state (cf. Fig. \ref{fig:HOSPS} (a)). This clearly shows the role of hole burning operations in inducing nonclassicality for odd orders. However, in case of even orders, the same operations are also observed to destroy the nonclassicality in the parent state. From Figs. \ref{fig:HOSPS} (b) and (c), it
is observed that  BS and KS do not show HOSPS  for the odd
values of $l$ even after application of state engineering 
operations. Additionally, HOSPS is not observed for the KS for even values of $l$, too. Consequently, the nonclassical feature witnessed through the HOSPS criterion in PAKS can be attributed solely to the hole burning process. 

%-------------------- Figure-4 ---------------------------------------------------
%----------------------------------------------------------------------------------
%---------------------------- Figure 5 ----------------------------
\begin{figure}
\centering{}
\includegraphics[scale=0.7]{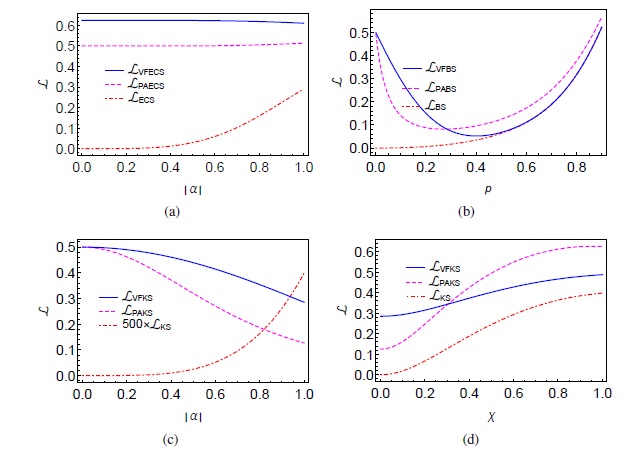}
\caption{\label{fig:Linear-Entropy}(Color online) Illustration of linear entropy for (a)
ECS, PAECS and VFECS, (b) BS, PABS and VFBS, (c) KS, PAKS and VFKS with $\alpha$ or $p$ for $\chi=0.02$.
(d) Dependence of nonclassicality in KS, PAKS and VFKS on $\chi$ for $\alpha=1$.}
\end{figure}
%----------------------------------------------------------------------------------

\section{Nonclassicality measure}

In 2005, a measure of nonclassicality
was proposed as entanglement potential, which is the amount of entanglement in two output ports of a beam splitter with the quantum state $\rho_{in}$ and vacuum $|0\rangle\langle0|$ sent through two input ports \cite{asboth2005computable}. The amount of entanglement  quantifies the amount of nonclassicality in the input quantum state as classical state can not generate entanglement in the output. The post beam splitter state
can be obtained as $\rho_{out}=U\left(\rho_{in}\otimes|0\rangle\langle0|\right)U^{\dagger}$ with $U=\exp[-iH\theta]$, where $H=(a^{\dagger}b+ab^{\dagger})/2$,
and $a^{\dagger}\,(a)$, $b^{\dagger}\,(b)$ are the creation (annihilation)
operators of the input modes. For example, considering quantum state (\ref{eq:fock-superposition}) and a vacuum state $|0\rangle$ as input states, we can write the analytic expression of the
two-mode output state as
\begin{equation}
|\phi\rangle=U\left(|\psi\rangle\otimes|0\rangle\right)\equiv U|\psi,0\rangle=\sum_{n=0}^{\infty}\,\frac{c_{n}}{2^{n/2}}\sum_{j=0}^{n}\sqrt{^{n}C_{j}}\,\,|j,\,n-j\rangle.\label{eq:inout_psi}
\end{equation}
We can measure the amount of entanglement in the output state to quantify the amount of input nonclassicality in $|\psi\rangle$. Here, we use linear entropy of single mode subsystem (obtained after tracing over the other subsystem) as entanglement potential. The linear entropy
for an arbitrary bipartite state $\rho_{AB}$ is defined
as \cite{wei2003maximal}
\begin{equation}
\mathcal{L}=1-{\rm Tr}\left(\rho_{B}^{2}\right),\label{eq:le}
\end{equation}
where $\rho_{B}$ is obtained by tracing over subsystem $A$.
We have obtained the analytic expressions of linear entropy for ECS, KS, BS
and corresponding engineered states and have reported them as Appendix A Eqs. (\ref{eq:LE-ECS})-(\ref{eq:LE-PAKS}). In general, significance of hole burning operations can be clearly established through corresponding results shown in Fig. \ref{fig:Linear-Entropy}. Specifically, one can clearly see the amount of nonclassciality (revealed through the amount of entanglement it can generate at a beam splitter) increases due to these operations. 

\begin{figure}
\centering{}\includegraphics[scale=0.75]{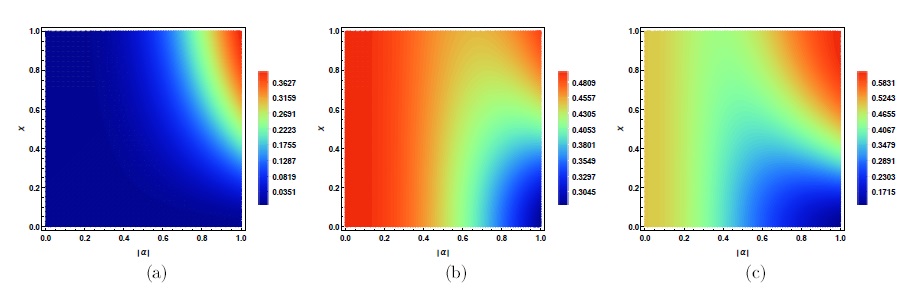}
\caption{\label{fig:LE}(Color online) Illustration of linear entropy  for (a) KS (b) VFKS (c) PAKS.}
\end{figure}

From Figs. \ref{fig:Linear-Entropy} (a) and (c), it can be observed that
vacuum filtered ECS and KS are more nonclassical than corresponding photon added
counterparts. However, in case of BS and its engineered states, it is observed that only up to a certain value of $p$ VFBS is more nonclassical than PABS (cf. Fig. \ref{fig:Linear-Entropy} (a)). In fact, the amount of additional nonclassicality induced due to filtration decreases with $p$ and eventually becomes zero (i.e., the amount of nonclassicality of VFBS becomes equal to that of BS as far as linear entropy is considered as a measure of nonclassicality).  It is interesting to observe the effect of Kerr coupling parameter $\chi$ on the amount of nonclassicality induced due to nonlinearity. It is observed that for small (relatively large) values of $\chi$ nonclassicality present in VFKS (PAKS) is more than that in PAKS (VFKS) (cf. Fig. \ref{fig:Linear-Entropy} (d)). This dependence is more clearly visible in Fig. \ref{fig:LE}, where one can observe strong nonclassicality in PAKS and VFKS  (KS) favor (favors) smaller (higher) values of $\alpha$ and large $\chi$. 

%------------------------ Figure-6----------------------------------------------
%-------------------------------------------------------------------------------

\section{Conclusion \label{sec:Conclusion}}

In summary, this article is focused on the comparison of the effects
of two processes (vacuum state filtration and single photon addition)
used in quantum state engineering to burn hole at vacuum as far as
the higher-order nonclassical properties of the quantum states prepared
using these two processes are concerned. Specifically, various quantum state engineering processes
for burning holes at vacuum lead to different  $\sum_{m=1}c_{m}|m\rangle$
as far as the values of $c_{m}$s are concerned (even when the parent
state is the same). To study its significance in nonclassical properties of the engineered states, we considered a small set of finite and infinite dimensional quantum states (namely, ECS, BS, and KS). This provided us a set of six engineered quantum states, namely VFECS,
PAECS, VFBS, PABS, VFKS, and PAKS and three parent states for our analysis. This set of engineered quantum
states can have a great importance in quantum information processing
and quantum optics as they are found to be highly nonclassical. Especially
when some exciting applications of their parent states are already investigated
in the context of continuous variable quantum information processing
and/or quantum optics. The present study also addresses the significance
of these hole burning processes in inducing (enhancing) particular nonclassical features in the large set of engineered and parent quantum states. 

The general expressions
for moments of the set of states are reported in the compact analytic form, which are used here to investigate nonclassical features of these states using
a set of criteria of higher-order nonclassicality (e.g., criteria
of HOA, HOS and HOSPS). The obtained expressions can be further used to study other moment-based criteria of nonclassicality. The hole burning operations are found to be extremely relevant as the states studied here are found to be highly
nonclassical when quantified through a measure of nonclassicality (entanglement potential). 
%A summary of the comparison performed here is given in Table \ref{tab:Observation-of-lower-}.
 In brief, both the vacuum filtration and photon addition operations can be ascribed as antibunching inducing operations in KS and ECS while antibunching enhancing operations for BS. As far as HOS is concerned no such advantage of these operations is visible as these operations fail to induce squeezing in ECS and often decrease the amount of squeezing present in the parent state. Additionally, the operations are successful in inducing HOSPS in KS and enhances this feature in the rest of the parent states. The relevance of higher-order nonclassicality in the context of the present study can be understood from the fact that these hole burning operations show an increase in the depth of HOA witness and decrease in the amount of HOS with order. While in case of HOSPS even orders show nonclassicality whereas odd orders fail to detect it. Finally, the measure of nonclassicality reveals vacuum filtration as a more powerful tool than photon addition for enhancing nonclassicality in the parent state, but photon addition is observed to be advantageous in some specific cases.

Based on our present investigation and the discussion above, we conclude
 the paper with the hope that the
methods used here will be helpful in further theoretical studies on
nonclassical properties of engineered quantum states (both finite
and infinite dimensional), and the results reported here
will be used in the experimental quantum information processing and/or
quantum optics.

\section*{Acknowledgment}

AP, SB and VN acknowledge the support from Interdisciplinary Cyber   Physical   Systems   (ICPS)   programme   of   the Department  of  Science  and  Technology (DST),  India,  Grant No.:DST/ICPS/QuST/Theme-1/2019/6. KT acknowledges the financial support from the Operational Programme Research, Development and Education - European Regional Development Fund project no. CZ.02.1.01/0.0/0.0/16 019/0000754 of the Ministry of Education, Youth and Sports of the Czech Republic.

\bibliographystyle{apsrev4-1}
\bibliography{biblio}

%-----------------------------APPENDIX--------------------------------------------
%\newpage
\appendix
\section*{Appendix: A}  % use *-form to suppress numbering

\renewcommand{\theequation}{A-\arabic{equation}}    
  % redefine the command that creates the equation no.    
  \setcounter{equation}{0}  % reset counter     
  
\subsection*{Analytic expression for linear entropy}

Analytical expression for linear entropy of ECS 
\begin{equation}
\begin{array}{lcl}
\mathcal{L}_{{\rm ECS}} & = & 1-\frac{\exp\left[-2\mid\alpha\mid^{2}\right]}{4\left(1+\exp\left[-2\mid\alpha\mid^{2}\right]\right)^{2}}\sum\limits _{n,m,r=0}^{\infty}\frac{\mid\alpha\mid^{2n+2r}\left(1+\left(-1\right)^{n}\right)\left(1+\left(-1\right)^{m}\right)\left(1+\left(-1\right)^{r}\right)\left(1+\left(-1\right)^{n+r-m}\right)}{n!r!} \sum\limits _{k_{1}=0}^{n}{n \choose k_{1}}{r \choose r+k_{1}-m}\left(\frac{1}{2}\right)^{n+r},
\end{array}\label{eq:LE-ECS}
\end{equation}
VFECS
\begin{equation}
\begin{array}{l}
\mathcal{L}_{{\rm VFECS}}=1-\left(N_{{\rm VFECS}}\right)^{4}\sum\limits _{n,m,r=1}^{\infty}\frac{\mid\alpha\mid^{2n+2r}\left(1+\left(-1\right)^{n}\right)\left(1+\left(-1\right)^{m}\right)\left(1+\left(-1\right)^{r}\right)\left(1+\left(-1\right)^{n+r-m}\right)}{n!r!}\sum\limits _{k_{1}=0}^{n}{n \choose k_{1}}{r \choose r+k_{1}-m}\left(\frac{1}{2}\right)^{n+r},\end{array}\label{eq:LE-VFECS}
\end{equation}
and PAECS
\begin{equation}
\begin{array}{lcl}
\mathcal{L}_{{\rm PAECS}} & = & 1-\left(N_{PAECS}\right)^{4}\sum\limits _{n,m,r=0}^{\infty}\frac{\mid\alpha\mid^{2n+2r}\left(1+\left(-1\right)^{n}\right)\left(1+\left(-1\right)^{m}\right)\left(1+\left(-1\right)^{r}\right)\left(1+\left(-1\right)^{n+r-m}\right)}{n!r!}\\
 & \times & \sum\limits _{k_{1}=0}^{n+1}{n+1 \choose k_{1}}{r+1 \choose r+k_{1}-m}\left(\frac{1}{2}\right)^{n+r+2}\left(m+1\right)\left(n-m+r+1\right).
\end{array}\label{eq:LE-PAECS}
\end{equation}
Similarly, analytical expression for linear entropy of BS 
\begin{equation}
\begin{array}{lcl}
\mathcal{L}_{{\rm BS}} & = & 1-\sum\limits _{n,m,r=0}^{M}\frac{1}{n!r!}\left[\frac{(M!)^{4}p^{2(n+r)}(1-p)^{4M-2n-2r}}{(M-n)!(M-m)!(M-r)!(M-n-r+m)!}\right]^{1/2} \sum\limits _{k_{1}=0}^{n}{n \choose k_{1}}{r \choose r+k_{1}-m}\left(\frac{1}{2}\right)^{n+r},
\end{array}\label{eq:LE-BS}
\end{equation}
VFBS
\begin{equation}
\begin{array}{lcl}
\mathcal{L}_{{\rm VFBS}} & = & 1-\left(N_{{\rm VFBS}}\right)^{4}\sum\limits _{n,m,r=1}^{M}\frac{1}{n!r!}\left[\frac{(M!)^{4}p^{2(n+r)}(1-p)^{4M-2n-2r}}{(M-n)!(M-m)!(M-r)!(M-n-r+m)!}\right]^{1/2} \sum\limits _{k_{1}=0}^{n}{n \choose k_{1}}{r \choose r+k_{1}-m}\left(\frac{1}{2}\right)^{n+r},
\end{array}\label{eq:LE-VFBS}
\end{equation}
and PABS 
\begin{equation}
\begin{array}{lcl}
\mathcal{L}_{{\rm PABS}} & = & 1-\left(N_{{\rm PABS}}\right)^{4}\sum\limits _{n,m,r=0}^{M}\frac{1}{n!r!}\left[\frac{(M!)^{4}p^{2(n+r)}(1-p)^{4M-2n-2r}}{(M-n)!(M-m)!(M-r)!(M-n-r+m)!}\right]^{1/2}\\
 & \times & \sum\limits _{k_{1}=0}^{n+1}{n+1 \choose k_{1}}{r+1 \choose r+k_{1}-m}\left(\frac{1}{2}\right)^{n+r+2}\left(m+1\right)\left(n-m+r+1\right)
\end{array}\label{eq:LE-PABS}
\end{equation}
are obtained. Finally, analytical expression for linear entropy of KS, VFKS, PAKS can be given
as
\begin{equation}
\begin{array}{lcl}
\mathcal{L}_{{\rm KS}} & = & 1-\sum\limits _{n,m,r=0}^{\infty}\frac{\left|\alpha\right|^{2n+2r}\exp\left[-2\mid\alpha\mid^{2}\right]}{n!r!}\exp[\iota\chi(m(m-1)-n(n-1)-r(r-1)\\
 & + & (n-m+r)(n-m+r-1))]\sum\limits _{k_{1}=0}^{n}{n \choose k_{1}}{r \choose r+k_{1}-m}\left(\frac{1}{2}\right)^{n+r},
\end{array}\label{eq:LS-KS}
\end{equation}
\begin{equation}
\begin{array}{lcl}
\mathcal{L}_{{\rm VFKS}} & = & 1-\left(N_{{\rm VFKS}}\right)^{4}\sum\limits _{n,m,r=1}^{\infty}\frac{\left|\alpha\right|^{2n+2r}}{n!r!}\exp[\iota\chi(m(m-1)-n(n-1)-r(r-1)\\
 & + & (n-m+r)(n-m+r-1))]\sum\limits _{k_{1}=0}^{n}{n \choose k_{1}}{r \choose r+k_{1}-m}\left(\frac{1}{2}\right)^{n+r},
\end{array}\label{eq:LE-VFKS}
\end{equation}
and
\begin{equation}
\begin{array}{lcl}
\mathcal{L}_{{\rm PAKS}} & = & 1-\left(N_{{\rm PAKS}}\right)^{4}\sum\limits _{n,m,r=0}^{\infty}\frac{\mid\alpha\mid^{2n+2r}}{n!r!}\exp[\iota\chi(m(m-1)-n(n-1)-r(r-1)\\
 & + & (n-m+r)(n-m+r-1))]\sum\limits _{k_{1}=0}^{n+1}{n+1 \choose k_{1}}{r+1 \choose r+k_{1}-m}\left(\frac{1}{2}\right)^{n+r+2}\left(m+1\right)\left(n-m+r+1\right),
\end{array}\label{eq:LE-PAKS}
\end{equation}
respectively.

\end{document}